\input epsf.tex
 \documentclass[12pt,twoside]{article}

\newcommand{\bmat}{\left(\begin{array}}
\newcommand{\emat}{\end{array}\right)}
\def\NPB{Nucl. Phys. B}

\def\yzero{\smash{\hbox{$y\kern-4pt\raise1pt\hbox{${}^\circ$}$}}}

\def\a{\alpha}
\def\b{\beta}
\def\g{\gamma}

\def\ep{\epsilon}

\def\OR{\Omega {\cal R}}

\def\ent{{\bf Z}}

\def\OR{\Omega {\cal R}}

\def\om{\omega}

\def\-{\hphantom{-}}
\def\ov{\overline}
\def\s2{\frac{1}{\sqrt2}}

\def\beq{\begin{equation}}
\def\eeq{\end{equation}}
\def\beqa{\begin{eqnarray}}
\def\eeqa{\end{eqnarray}}
\def\ba{\begin{array}}
\def\ea{\end{array}}

\def\IF{\relax{\rm I\kern-.18em F}}
\def\II{\relax{\rm I\kern-.18em I}}
\def\IP{\relax{\rm I\kern-.18em P}}
\def\IC{\relax\hbox{\kern.25em$\inbar\kern-.3em{\rm C}$}}
\def\IR{\relax{\rm I\kern-.18em R}}

\def\cp{{\cal P}}

\def\Dsl{\,\raise.15ex\hbox{/}\mkern-13.5mu D} 
\def\IZ{Z\kern-.4em  Z}

 \def\cp#1{\relax\ifmmode {\IP\kern-2pt{}_{#1}}\else $\IP\kern-2pt{}_{#1}$\=fi}

\catcode`\@=11
\newdimen\@rotdimen
\newbox\@rotbox

\def\@vspec#1{\special{ps:#1}}
\def\@rotstart#1{\@vspec{gsave currentpoint currentpoint translate
   #1 neg exch neg exch translate}}
\def\@rotfinish{\@vspec{currentpoint grestore moveto}}
\def\@rotr#1{\@rotdimen=\ht#1\advance\@rotdimen by\dp#1%
   \hbox to\@rotdimen{\hskip\ht#1\vbox to\wd#1{\@rotstart{90 rotate}%
   \box#1\vss}\hss}\@rotfinish}
\def\@rotl#1{\@rotdimen=\ht#1\advance\@rotdimen by\dp#1%
   \hbox to\@rotdimen{\vbox to\wd#1{\vskip\wd#1\@rotstart{270 rotate}%
   \box#1\vss}\hss}\@rotfinish}%
\def\@rotu#1{\@rotdimen=\ht#1\advance\@rotdimen by\dp#1%
   \hbox to\wd#1{\hskip\wd#1\vbox to\@rotdimen{\vskip\@rotdimen
   \@rotstart{-1 dup scale}\box#1\vss}\hss}\@rotfinish}%
\def\@rotf#1{\hbox to\wd#1{\hskip\wd#1\@rotstart{-1 1 scale}%
   \box#1\hss}\@rotfinish}%
\def\rotate{\@ifnextchar[{\@rotate}{\@rotate[l]}}
\def\@rotate[#1]#2{\setbox\@rotbox=\hbox{#2}\@nameuse{@rot#1}\@rotbox}

\catcode`\@=12

\begin{document}

\makeatletter
\@addtoreset{equation}{section} \makeatother
\renewcommand{\theequation}{\thesection.\arabic{equation}}
\pagestyle{empty}
\pagestyle{plain}
\vspace{0.5cm}
\setcounter{footnote}{0}

\begin{center}
{\LARGE{\bf Standard Model Building 
from Intersecting D-branes }}
\\[7mm]
{\Large{{  Christos ~Kokorelis } }
\\[2mm]}
\small
{\em Nuclear and Particle Physics Sector, Univ. of Athens, 
 GR-15771, Athens, Greece}
\\[0.3em]
{\em and}\\[0.3em]
{\em 
 Institute of Nuclear Physics, N.C.S.R. Demokritos, GR-15310, Athens, Greece}
\end{center}
\vspace{1mm}


\begin{center}
{\small \bf ABSTRACT}
\end{center}

We provide a general overview of the current state of the art in 
four dimensional three generation model building proposals - using intersecting 
D-brane toroidal compactifications [without fluxes]
of IIA, IIB string theories - which have only the SM at low energy. In this 
context,  
we focus on these model building directions, where non-supersymmetric 
constructions - based on the existence of the gauge group structure 
$SU(3)_c \times SU(2)_L \times U(1)_Y$, Pati-Salam
 $SU(4)_C \times SU(2)_L \times SU(2)_R$,  
SU(5) and flipped SU(5) GUTS - appear at the string scale $M_s$. These model building 
attempts 
are based on four dimensional 
compactifications that use orientifolds of either IIA theory with D6-branes wrapping 
on $T^6$, $T^6/Z_3$ and recently on $T^6/Z_3 \times Z_3$ or of IIB theory with D5-branes 
wrapping on $T^4 \times C/Z_N$.  
Models with D5-branes 
are compatible with the large extra dimension scenario and a low string scale that 
could be at the TeV; thus there is no gauge hierarchy 
problem in the Higgs sector.
In the case of 
flipped SU(5) GUTS - coming from $T^6/Z_3$ - the special build up structure 
of the models accommodates naturally a see-saw mechanism and a new solution to the
doublet-triplet splitting problem. Baryon number is a gauged symmetry and thus proton 
is naturally stable only in models with D5 branes or in models with D6-branes wrapping toroidal
orientifolds of type IIA.  In the rest of the constructions, proton is stable due to the 
existence of a high $M_s$ that suppress gauge mediating baryon violating 
couplings.  
Extra beyond the SM gauge group U(1)'s that may be present at $M_s$ and have 
non-zero couplings to RR fields,
 become 
massive by the existence of a generalized Green-Schwarz mechanism and by the 
existence of either
tachyonic singlets receiving a vev or gauge neutral singlets becoming massless at N=1 
supersymmetric intersections.  
Models with intersecting D-branes accommodate
naturally right handed neutrinos. Finally, we present new RR tadpole solutions
 for the 5- and 6- stack toroidal orientifold models of type IIA which have only the Standard Model
with right handed neutrinos at low energy.


\newpage

\section{Prelimiraries - D6/D5-brane model building with only
the SM at Low Energy}

At present string theory is our only candidate  
theory that could unify gravity with the rest of the fundamental interactions 
in a perturbative 
framework.  In this respect, in the last three years, string constructions based 
on D-branes intersecting at angles - intersecting brane worlds (IBW's) for 
short - received a lot of attention \cite{iba} -  
\cite{kokosnew}, as on these constructions, 
for the first time in string theory, it became 
possible \footnote{See also \cite{rev} for other reviews on the subject} 
to construct four dimensional non-supersymmetric (non-susy)
intersecting D6-brane 
models that have only the Standard model 
(SM) gauge group and chiral spectrum - with right handed neutrinos $\nu_R$'s 
- at low energies \cite{iba, kokos1, kokos2}. For N=1 supersymmetric model building
attempts in IBW's see for a partial list of references \cite{cve1, blu10}.
In IBW's chiral matter get localized as open strings stretching in the 
intersections between intersecting D-branes \cite{dou}.
Subsequently, the chiral spectrum of the model may be obtained by
solving simultaneously the intersection constraints coming from the
existence of the different sectors and the RR tadpole cancellation
conditions.

\begin{figure}
\begin{center}
\centering
\epsfysize=6cm
\leavevmode
\epsfbox{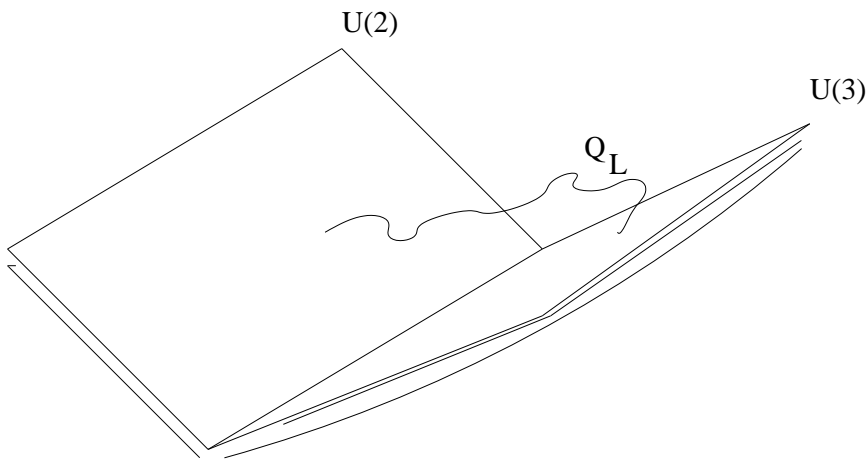}
\end{center}
\caption[]{\small
Open strings stretching between intersecting branes get identified as fermionic 
matter. A left handed quark makes its appearance.
 }
\label{figa1}
\end{figure}

Thus e.g. if a stack of parallel D-branes gets associated to 
the U(3) and another one to the U(2) gauge groups, the chiral fermion localized 
in their intersection represents a quark; see fig. (\ref{figa1}).  
Hence by considering $a$ stacks of D-brane configurations
with $N_a, a=1, \cdots, N$, parallel branes in each stack, we get the gauge group
configuration $U(N_1) \times U(N_2) \times \cdots \times U(N_a)$.  Each
$U(N_i)$ factor will also give rise to an $SU(N_i)$ charged under the
associated $U(1_i)$ gauge group factor that appears in the
decomposition $SU(N_a) \times U(1_a)$.

The initial constructions of IBW's exhibiting the SM at low energy,
were based on a background of D6-branes intersecting at angles in
4D toroidal orientifolds \cite{lu} of [T-dual to models with magnetic 
deformations \cite{ang}. See also \cite{pra}.] type
IIA theory and exhibit proton stability as baryon number is a gauged 
symmetry.
The primary common phenomenological characteristics of the four stack
D6-models of \cite{iba} and the five and six SM's of \cite{kokos1} and 
\cite{kokos2} respectively - emphasizing that there are 
no D6-brane models with only the SM at low energy using 
constructions with more than six stacks of D6-branes at the 
string scale $M_s$ -
are : \newline
$\bullet$  the prediction of the 
existence of the 
SM chiral spectrum together with $\nu_R$'s,
\newline
$\bullet$ 
proton stability; conservation of lepton number - the models admit Dirac terms for the 
neutrinos -  their masses appear as a result of the 
existence of particular Yukawa couplings associated with the breaking
of the chiral symmetry.\newline
Additionally, the SM's of \cite{kokos1} and 
\cite{kokos2} {\em exhibit a new phenomenon} - not found in the SM's of \cite{iba},
namely {\em the prediction of the existence of 
N=1 supersymmetric (SUSY) 
partners of 
$\nu_R$'s, the s$\nu_R$'s.} 
Thus even though the models of \cite{kokos1, kokos2}
are non-susy, they have particular N=1 SUSY partners, the s$\nu_R$'s. 
There are two ways that vacua with only the SM have been produced in IBW's : a) 
with all matter accommodated in bifundamental
representations (reps) \cite{iba, kokos1, kokos2} $(N, {\bar N})$ of $U(N_a) 
\times U(N_b)$. In this case, we have the clear advantage that all matter Yukawa's may be 
realized,  
 b) when some of the matter fields appear 
in symmetric and antisymmetric reps of $U(N_a)$ \cite{lust3, kokos4, kokosnew}. In this case  
some of the SM matter mass couplings may be missing.

Take for example the general picture with D${(3+l)}_a$-branes 
wrapping $l$-cycles with 
$(n^l_a, m^l_a)$, $l=1,2,3$ the wrapping numbers of a brane along each of 
the $(T^{2l})$
torus.
Thus we allow the six-torus to wrap factorized 2-cycles, so we can
unwrap the l-cycle into products of 1-cycles.
The definition of the homology of the l-cycles as
\beq
[\Pi_a] =\ \prod_{i=1}^l(n^i_a [a_i] + m^i_a[b_i])
\label{homo1}
\eeq
The number of fermions $I_{ab}$ localized in an intersection is described by the intersection number $I_{ab} =[\Pi_a]  [\Pi_b]$. 
Major role in these constructions is played by the satisfaction of the so called
RR tadpole cancellation conditions (TCC), which in low energies is seen as the 
cancellation of cubic gauge anomalies. TCC's can be alternatively be expressed
as the cancellation of the RR charge in homology
\beq
\sum_a N_a [ \Pi_a ] = 0
\eeq
We note that while in N=1 susy constructions NSNS automatically cancel, in non-susy 
constructions 
uncancelled NS 
tadpoles remain; even though it 
 remains an open issue
whether or not they can be cancelled in higher orders of
perturbation theory. 
The NS tadpoles do not 
affect the low energy spectrum of the models but they rather imply 
the instability of the models, in the present order of perturbation 
theory, in a flat background. 

In the absence of a principle for selecting a particular string vacuum, 
we select to examine
whether or not it is possible to obtain non-susy models with just the SM  at low energy by 
examining different 4D string compactifications. The method that has been 
used \cite{iba, kokos1, kokos2} follows a bottom-up
approach. In particular we embed the localization of SM fermions at 
particular intersections - as it is seen in tables (\ref{spectrum7a}), 
(\ref{spectrum7b}), (\ref{spectrum7c}) - to different four dimensional (4D) 
compactifications. As we will see in sections 2 and 3, for constructions that involve a 
SM-like configuration 
at $M_s$ - e.g. $U(3) \times U(2) \times U(1)^n$, n= 2,3,4 - 
 these tables were embedded successfully to 4D models 
with D6-branes wrapping on IIA compactified on a toroidal orientifold 
\cite{iba, kokos1, kokos2} or to 4D models with
D5-branes wrapping $T^4 \times C/Z_N$ \cite{ibaD5, kokoD5}.

\begin{table}[htb] \footnotesize
\renewcommand{\arraystretch}{1}
\begin{center}
\begin{tabular}{|c|c|c|c|c|c|c|c|c|}
\hline
Matter Fields & & Intersection & $Q_a$ & $Q_b$ & $Q_c$ & $Q_d$ & Y
\\\hline
 $Q_L$ &  $(3, 2)$ & $I_{ab}$ & $1$ & $-1$ & $0$ & $0$ &  $1/6$ \\\hline
 $q_L$  &  $2(3, 2)$ & $I_{a b^{\ast}}$ &
$1$ & $1$ & $0$ & $0$  & $1/6$  \\\hline
 $U_R$ & $3({\bar 3}, 1)$ & $I_{ac}$ &
$-1$ & $0$ & $1$ & $0$ & $-2/3$ \\\hline
 $D_R$ &   $3({\bar 3}, 1)$  &  $I_{a c^{\ast}} $ &
$-1$ & $0$ & $-1$ & $0$ & $1/3$ \\\hline
$L$ &   $3(1, 2)$  &  $I_{bd} $ &
$0$ & $-1$ & $0$ & $1$ & $-1/2$  \\\hline
$N_R$ &   $3(1, 1)$  &  $I_{cd} $ &
$0$ & $0$ & $1$ & $-1$ & $-1/2$  \\\hline
$E_R$ &   $3(1, 1)$  &  $I_{cd^{\star}}$ &
$0$ & $0$ & $-1$ & $-1$ & $0$  \\\hline
\end{tabular}
\end{center}
\caption{\small Low energy fermionic spectrum of the
four stack
string scale
$SU(3)_C \otimes
SU(2)_L \otimes U(1)_a \otimes U(1)_b \otimes U(1)_c
\otimes U(1)_d $, D5-brane model together with its
$U(1)$ charges.
\label{spectrum7a}}
\end{table}

\begin{table}[htb] \footnotesize
\renewcommand{\arraystretch}{0.8}
\begin{center}
\begin{tabular}{|c|c|c|c|c|c|c|c|c|}
\hline
Matter Fields & & Intersection & $Q_a$ & $Q_b$ & $Q_c$ & $Q_d$ & $Q_e$& Y
\\\hline
 $Q_L$ &  $(3, 2)$ & $I_{ab}$ & $1$ & $-1$ & $0$ & $0$ & $0$& $1/6$ \\\hline
 $q_L$  &  $2(3, 2)$ & $I_{a b^{\ast}}$ &
$1$ & $1$ & $0$ & $0$  & $0$ & $1/6$  \\\hline
 $U_R$ & $3({\bar 3}, 1)$ & $I_{ac}$ &
$-1$ & $0$ & $1$ & $0$ & $0$ & $-2/3$ \\\hline
 $D_R$ &   $3({\bar 3}, 1)$  &  $I_{a c^{\ast}} $ &
$-1$ & $0$ & $-1$ & $0$ & $0$ & $1/3$ \\\hline
$L$ &   $2(1, 2)$  &  $I_{bd} $ &
$0$ & $-1$ & $0$ & $1$ & $0$ & $-1/2$  \\\hline
$l_L$ &   $(1, 2)$  &  $I_{b e} $ &
$0$ & $-1$ & $0$ & $0$ & $1$ & $-1/2$  \\\hline
$N_R$ &   $2(1, 1)$  &  $I_{cd}$ &
$0$ & $0$ & $1$ & $-1$ & $0$ & $0$  \\\hline
$E_R$ &   $2(1, 1)$  &  $I_{c d^{\ast}} $ &
$0$ & $0$ & $-1$ & $-1$ & $0$ & $1$   \\\hline
  $\nu_R$ &   $(1, 1)$  &  $I_{c e} $ &
$0$ & $0$ & $1$ & $0$ & $-1$ & $0$ \\\hline
$e_R$ &   $(1, 1)$  &  $I_{c e^{\ast}} $ &
$0$ & $0$ & $-1$ & $0$ & $-1$  & $1$ \\\hline
\hline
\end{tabular}
\end{center}
\caption{\small Low energy fermionic spectrum of the five stack
string scale
$SU(3)_C \otimes
SU(2)_L \otimes U(1)_a \otimes U(1)_b \otimes U(1)_c
\otimes U(1)_d \otimes U(1)_e $, D5-brane model together with its
$U(1)$ charges.
\label{spectrum7b}}
\end{table}

\begin{table}[htb] \footnotesize
\renewcommand{\arraystretch}{0.8}
\begin{center}
\begin{tabular}{|c|c|c|c|c|c|c|c|c|c|}
\hline
Matter Fields & & Intersection & $Q_a$ & $Q_b$ & $Q_c$
& $Q_d$ & $Q_e$ & $Q_f$ & Y \\\hline
 $Q_L$ &  $(3, 2)$ & $I_{ab}$ &
 $1$ & $-1$ & $0$ & $0$ & $0$ & $0$ & $1/6$ \\\hline
 $q_L$  & $2(3, 2)$ & $I_{a b^{\ast}}$ &
$1$ & $1$ & $0$ & $0$  & $0$ & $0$ & $1/6$  \\\hline
 $U_R$ & $3({\bar 3}, 1)$ & $I_{ac}$ &
$-1$ & $0$ & $1$ & $0$ & $0$ & $0$ & $-2/3$ \\\hline
 $D_R$ &   $3({\bar 3}, 1)$  &  $I_{a c^{\ast}} $ &
$-1$ & $0$ & $-1$ & $0$ & $0$ & $0$ & $1/3$ \\\hline
$L^1$ &   $(1, 2)$  &  $I_{bd} $ &
$0$ & $-1$ & $0$ & $1$ & $0$ & $0$ & $-1/2$  \\\hline
$L^2$ &   $(1, 2)$  &  $I_{be} $ &
$0$ & $-1$ & $0$ & $0$ & $1$ & $0$ & $-1/2$  \\\hline
$L^3$ &   $(1, 2)$  &  $I_{bf} $ &
$0$ & $-1$ & $0$ & $0$ & $0$ & $1$ &$-1/2$  \\\hline
$N_R^1$ &   $(1, 1)$  &  $I_{cd}$ &
$0$ & $0$ & $1$ & $-1$ & $0$ & $0$ & $0$ \\\hline
$E_R^1$ &   $(1, 1)$  &  $I_{c d^{\ast}} $ &
$0$ & $0$ & $-1$ & $-1$ & $0$ & $0$ & $1$  \\\hline
$N_R^2$ &   $(1, 1)$  &  $I_{ce}$ &
$0$ & $0$ & $1$ & $0$ & $-1$ & $0$ & $0$ \\\hline
$E_R^2$ &   $(1, 1)$  &  $I_{c e^{\ast}} $ &
$0$ & $0$ & $-1$ & $0$ & $-1$ & $0$ & $1$ \\\hline
$N_R^3$ &   $(1, 1)$  &  $I_{cf}$ &
$0$ & $0$ & $1$ & $0$ & $0$ & $-1$ & $0$ \\\hline
$E_R^3$ &   $(1, 1)$  &  $I_{c f^{\ast}} $ &
$0$ & $0$ & $-1$ & $0$ & $0$ & $-1$ & $1$ \\\hline
\end{tabular}
\end{center}
\caption{
\small Low energy fermionic spectrum of the six stack
string scale
$SU(3)_C \otimes
SU(2)_L \otimes U(1)_a \otimes U(1)_b \otimes U(1)_c
\otimes U(1)_d \otimes U(1)_e \otimes U(1)_f$, D5-brane model together with its
$U(1)$ charges.
\label{spectrum7c}}
\end{table}

The hypercharge operator of the SM's appearing in these tables 
is defined as a linear combination
of the $U(1)_a$, $U(1)_c$, $U(1)_d$, $U(1)_e$, $U(1)_f$
gauge groups for the four-, five-, six-stack models respectively:
\beq
Y_{[Table \ \ref{spectrum7a}] }  = \frac{1}{6}U(1)_{a}- \frac{1}{2} U(1)_c -
\frac{1}{2} U(1)_d\;.
\label{hyper1230}
\eeq
\beq
Y_{[Table \ \ref{spectrum7b}] } = \frac{1}{6}U(1)_{a}- \frac{1}{2} U(1)_c -
\frac{1}{2} U(1)_d -
 \frac{1}{2} U(1)_e \;.
\label{hyper1231}
\eeq
\beq
Y_{[Table \ \ref{spectrum7c}] } = \frac{1}{6}U(1)_{a}- \frac{1}{2} U(1)_c -
\frac{1}{2} U(1)_d -
 \frac{1}{2} U(1)_e - \frac{1}{2} U(1)_f \;.
\label{hyper1232}
\eeq

Moreover the first non-SUSY constructions of 
string GUTS
which have only the SM at low energy \footnote{
These models were also based
on the intersecting D6-backgrounds of \cite{lu}.}, were
constructed in \cite{kokos5}, based on the Pati-Salam structure 
$SU(4)_C \times SU(2)_L \times SU(2)_R$ at $M_s$. These models may be described in 
section 3.

\section{Only the SM at low energy from models compatible with the large extra 
dimension scenario}

In this class of models \cite{ibaD5} the general picture 
involves D$5_a$-branes wrapping 1-cycles
$(n^i_a, m^i_a)$, $i=1,2$ along each of the ith-$T^2$
torus of the factorized $T^4$ torus, namely
$T^4 = T^2 \times T^2$.
Thus we allow the four-torus to wrap factorized 2-cycles, so we can
unwrap the 2-cycle into products of two 1-cycles, one for each $T^2$.
The definition of the homology of the 2-cycles as
\beq
[\Pi_a] =\ \prod_{i=1}^2(n^i_a [a_i] + m^i_a[b_i])
\label{homo1}
\eeq
defines consequently the 2-cycle of the orientifold images as
\beq
[\Pi_{a^{\star}}] =\ \prod_{i=1}^2(n^i_a [a_i] - m^i_a[b_i]).
\label{homo2}
\eeq
We note that because of the $\Omega {\cal R}$ symmetry
 each D$5_a$-brane
1-cycle, must be accompanied by its $\Omega {\cal R}$
orientifold image
partner $(n^i_a, -m^i_a)$; $n, m \in Z$.
In addition, because of
the presence of discrete NS B-flux \cite{lu},
the tori involved are not
orthogonal but tilted. Hence, 
the wrapping numbers become the effective tilted wrapping numbers,
\beq
(n^i, m ={\tilde m}^i + b^i \cdot {n^i}/2);\;
n,\;{\tilde m}\;\in\;Z, \  b^i = 0, 1/2,
\label{na2}
\eeq
where semi-integer values are allowed for the m-wrappings.

Let us discuss the effect
of the orbifold action on the open string sectors.
The $\ent_N$ orbifold
twist in the third complex dimension is
generated by the twist vector
$v = \frac{ 1}{N} (0,0,-2,0)$, which is fixed
by the requirements of modular invariance and for
the variety to be spin. Subsequently the $\ent_N$ action is
embedded in the $U(N_a)$ degrees of freedom emanating from the
$a^{th}$ stack of D5-branes, through the unitary matrix in the
form
\beq
\g_{\om,a} = {\it diag} \left( {\bf 1}_{N_a^0},\ \alpha {\bf 1}_{N_a^1},
\ldots,\ \alpha^{N-1} {\bf 1}_{N_a^{N-1}} \right),
\label{Chan}
\eeq
with $\sum_{i=0}^{N-1} N_a^i = N_a$ and
$\a \equiv {\rm exp} (2\pi i/N)$.
In the presence of
$\OR$ orientifold action, we include
sectors where
its brane is accompanied by is orientifold image.
Lets us denote the $\OR$ image of the brane D5$_a$-brane
by D5$_{a^{\star}}$.
Hence if the D5$_a$-brane is described by the matrix
\beqa
& (n_a^1,m_a^1) \otimes (n_a^2,m_a^2) \nonumber \\
& \g_{\om,a} = {\it diag} \left( {\bf 1}_{N_a^0},\
\a {\bf 1}_{N_a^1},
\ldots,\ \a^{N-1} {\bf 1}_{N_a^{N-1}} \right),
\label{brana_a}
\eeqa
the D5$_{a^*}$ is given by
\beqa
& (n_a^1,-m_a^1) \otimes (n_a^2,-m_a^2) \nonumber \\
& \g_{\om,a^*} = {\it diag} \left( {\bf 1}_{N_a^0},\ \a^{N-1}
{\bf 1}_{N_a^{1}}, \ldots,\ \a {\bf 1}_{N_a^{N-1}} \right),
\label{brana_a*}
\eeqa

The RR tadpoles for the D5-branes at the $Z_N$ orbifold singularity are given 
by \cite{ibaD5}
{\beq
\begin{array}{l}
c_k^2 \ \sum_a n_a^1 n_a^2 \
\left({\rm Tr} \gamma_{k,a} + {\rm Tr} \gamma_{k,a^*} \right)
= 16 \ {\rm sin} \left(\frac{\pi k}{N} \right) \\
c_k^2 \ \sum_a m_a^1 m_a^2 \
\left({\rm Tr} \gamma_{k,a} + {\rm Tr} \gamma_{k,a^*} \right) = 0\\
c_k^2 \ \sum_a n_a^1 m_a^2 \
\left({\rm Tr} \gamma_{k,a} - {\rm Tr} \gamma_{k,a^*} \right) = 0\\
c_k^2 \ \sum_a m_a^1 n_a^2 \
\left({\rm Tr} \gamma_{k,a} - {\rm Tr} \gamma_{k,a^*} \right) = 0
\label{tadpoleO5b}
\end{array}
\eeq}
with $c_k^2= sin((2\pi k)/N)$.
The presence of a non-zero term in the first tadpole condition should be
interpreted as a negative RR charge induced by the presence of an O5-plane.
We note that the first of
twisted tadpole conditions can be also written as
\beq
\sum_a{n_a^1 n_a^2 \left({\rm Tr} \gamma_{2k,a}
+ {\rm Tr} \gamma_{2k,a^*}\right)} =
{16 \over {\alpha^k + \alpha^{-k}}}.
\label{decomp}
\eeq

\begin{figure}
\begin{center}
\centering
\epsfysize=6cm
\leavevmode
\epsfbox{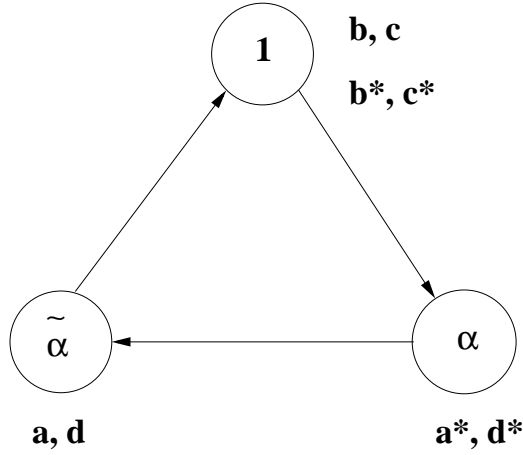}
\end{center}
\caption[]{\small
Assignment of SM embedding in
configurations of four stacks of D5 branes depicted by the `reflected'
$Z_3$ quiver diagrams. At low energy we get only the SM.
Note that ${\tilde \alpha} = \alpha^{-1}$. 
 }
\label{protifig}
\end{figure}

In this subsection we examine the derivation \cite{kokoD5} of exactly
the SM at
low energies from the embedding of the four stack SM structure of table (1)
in a $Z_3$ quiver of $Q1$-type seen in figure (\ref{protifig}).

\begin{table}[htb]\footnotesize
\renewcommand{\arraystretch}{1.4}
\begin{center}
\begin{tabular}{||c||c|c|c||}
\hline
\hline
$N_i$ & $(n_i^1, m_i^1)$ & $(n_i^2, m_i^2)$ & $(n_i^3, m_i^3)$\\
\hline\hline
 $N_a=3$ & $(n_a^1, \epsilon {\tilde \ep}\b^1)$  &
$(3,  \frac{1}{2}{\tilde \epsilon} \epsilon )$ & $1_3$  \\
\hline
$N_b=2$  & $(1/\b_1, 0)$ & $(1, \frac{1}{2}{\epsilon}{\tilde \ep})$ &
$\alpha^2 {\bf 1}_2$ \\
\hline
$N_c=1$ & $(1/\b_1, 0)$ &   $(0, {\epsilon{\tilde \ep}})$  &
$\alpha$ \\
\hline
$N_d=1$ & $(n_d^1, 3 \epsilon \b^1)$ &  $({\tilde \ep},  - \frac{1
}{2}\epsilon)$
  & $1$  \\\hline
$N_h$ & $(\epsilon_h/ \b^1, 0)$ &
$(2, 0 )$
  & $1_{N_h}$
\\\hline
\end{tabular}
\end{center}
\caption{\small General tadpole solutions for the four-stack
$Q1$-type quiver of intersecting
D5-branes, giving rise to exactly the
standard model gauge group and observable chiral spectrum
at low energies.
The solutions depend
on two integer parameters,
$n_a^1$, $n_d^1$,
the NS-background $\beta^1= \ 1-b_i$, which is associated to
the presence of the
NS B-field by $b_i =0,\ 1/2$.
 and
the phase parameters $\epsilon = {\tilde \epsilon}=\pm 1$, as well as the
CP phase $\alpha$.
\label{spectrum1oiel}}
\end{table}

The solutions satisfying simultaneously the
intersection constraints and the
cancellation of the RR twisted crosscap tadpole cancellation
constraints
are given in parametric form in table (\ref{spectrum1oiel}).
The multiparameter RR
tadpole solutions appearing in table (\ref{spectrum1oiel})
represent deformations of
the D5-brane branes, of table (1), intersecting at angles,
 within the same homology class of the
factorizable two-cycles.
The
solutions of table (\ref{spectrum1oiel}) satisfy all tadpole
equations, in (\ref{tadpoleO5b}), but the
first.
The latter reads :
\beq
9 n_a^1-\ \frac{1}{\b^1} +\ {\tilde \ep} \  n_d^1 + \
\frac{2\epsilon_h N_h}{\b^1} =\ -8.
\label{rampo}
\eeq

Note that we had added the presence of extra $N_h$ branes.
Their contribution to the RR tadpole conditions is best
described by placing them in the three-factorizable cycle
\beq
N_h \ (\epsilon_h/\b_1, 0)\ (2, 0)1_{N_h} \ .
\label{sda12el}
\eeq
The presence of an arbitrary number
of $N_h$ D5-branes, which give an extra $U(N_h)$ gauge group,
does not contribute to the rest of the tadpoles and
intersection constraints. Thus in terms of the
low energy theory no new chiral matter is generated and
only the SM spectrum appears. The analysis of the U(1) anomalies in the models
also shows that only the SM hypercharge survives massless at low energies. We will not 
present the details of this analysis as they can be found in detail in \cite{kokoD5}. 

In the present models, there are two dimensions transverse to the configuration space
that the D5-branes wrap. As the models are non-susy in order not to face the 
gauge hierarchy problem (GHP) in the Higgs sector of the models the string scale 
should be at the order
of the TeV. This may be realized by lowering the string scale while keeping the Planck scale 
large, by increasing the volume of the extra 2-dimensional space being transverse to the 
D5-branes. In this way, GHP is solved in consistency with the well known 
scenario of \cite{dva}.

\section{Building string vacua with D6-branes and only the SM at low energy}  
 
Historically, the first construction of models which have the SM at low 
energy -
with all matter in
bifundamental representations - appeared in \cite{iba}, while five and six stack models have been
studied in \cite{kokos1, kokos2} respectively.  
In this section we will describe the derivation of the 
SM at low energy using five (5) stacks of D6-branes at $M_s$. These models 
have been described before in \cite{kokos1} but in this section we will present a new
solution to the RR tadpoles not found in \cite{kokos1}. We will discuss in 
some detail the 5-stack SM solutions, while by the end of this section we 
will simply present the alternative RR tadpole cancellation 
solution to the 6-stack SM's that appear in \cite{kokos2}.

Next, we turn our attention to the construction of the standard models. It is based
on type I string with D9-branes compactified on a six-dimensional
orientifolded torus $T^6$, where internal background gauge fluxes on
the branes are turned on.  If we perform a T-duality transformation on
the $x^4$, $x^5$, $x^6$, directions the D9-branes with fluxes are
translated into D6-branes intersecting at angles. Note that the branes
are not parallel to the orientifold planes. Furthermore, we assume
that the D$6_a$-branes are wrapping 1-cycles $(n^i_a, m^i_a)$ along
each of the ith-$T^2$ torus of the factorized $T^6$ torus, namely $T^6
= T^2 \times T^2 \times T^2$.  That means that we allow our torus to
wrap factorized 3-cycles, that can unwrap into products of three
1-cycles, one for each $T^2$.  We define the homology of the 3-cycles
as
$[\Pi_a] = \prod_{i=1}^3 (n^i_a [a_i] + m^i_a[b_i])$
while we define the 3-cycle as in (\ref{homo1}) for the orientifold images as
$
[\Pi_{a^{\star}}] = \prod_{i=1}^3 (n^i_a [a_i] - m^i_a[b_i])$
In order to build the SM model structure a low energies, we consider
five stacks of D6-branes giving rise to their world-volume to an
initial gauge group $U(3) \times U(2) \times U(1) \times U(1)
\times U(1)$ or $SU(3) \times SU(2) \times U(1)_a \times U(1)_b
\times U(1)_c \times U(1)_d \times U(1)_e$ at the 
string scale.  Also, we consider the addition of NS
B-flux \cite{lu}, such that the tori involved are not orthogonal, and leading
to effective tilted wrapping numbers as in (\ref{na2}), thus 
allowing semi-integer values for the m-wrapping numbers.  

Because of the $\Omega {\cal R}$ symmetry, where $\Omega$ is the
worldvolume parity and $\cal R$ is the reflection on the T-dualized
coordinates, $T (\Omega {\cal R})T^{-1}= \Omega {\cal R}$
each D$6_a$-brane 1-cycle, must have its $\Omega {\cal R}$ image
partner $(n^i_a, -m^i_a)$.

 In the toroidal D6-models there
are a number of different sectors
contributing to the chiral spectrum.  To establish notation we denote the action of $\Omega
R$ on a sector $a, b$, by $a^{\star}, b^{\star}$, respectively.  We regognize the 
following sectors:

$\bullet$ The sector containing open strings stretching between
the D$6_a$ and D$6_b$ branes. Under the $\Omega R$ symmetry this
sector is mapped to its image.  The number of chiral fermions, namely $I_{ab}$,
transforms in the bifundamental representation $(N_a, {\bar
N}_a)$ of $U(N_a) \times U(N_b)$, and their multiplicity is given by 
\beq 
I_{ab} = [\Pi_a] \cdot [\Pi_b] = \left( n_a^1 m_b^1 - m_a^1
n_b^1\right)\left( n_a^2 m_b^2 - m_a^2 n_b^2 \right) \left(n_a^3 m_b^3
- m_a^3 n_b^3\right),
\label{ena3}
\eeq 
where $I_{ab}$ is the intersection number of the wrapped
cycles. Note from the sign of $I_{ab}$ intersection, we get 
the chirality of the fermions, where $I_{ab} > 0$ by convention denotes left handed
fermions. 

$\bullet$ The sector containing open string stretching between the brane $a$ and 
the orintifod image of the $b$-brane.   In this sector chiral
fermions transforming into the $(N_a, N_b)$ representation with
multiplicity given
\beq
I_{ab^{\star}} = [\Pi_a] \cdot [\Pi_{b^{\star}}] =\ -\left( n_a^1
m_b^1 + m_a^1 n_b^1\right)\left( n_a^2 m_b^2 + m_a^2 n_b^2 \right)
\left(n_a^3 m_b^3 + m_a^3 n_b^3\right).
\label{ena31}
\eeq
Similar conventions to  $I_{ab}$ hold for the sign of $I_{ab^{\star}}$.
 
$\bullet$ The sector containing open strings stretching between the brane $a$ and its 
orientifold image, namely $aa^{\star}$. In this sector the invariant intersections
contribute 8$m_a^1 m_a^2 m_a^3$ fermions in the antisymmetric
representation and the non-invariant intersections
add 4$ m_a^1 m_a^2 m_a^3 (n_a^1 n_a^2 n_a^3 -1)$
additional fermions in the symmetric and antisymmetric representation
of the $U(N_a)$ gauge group.

In general the calculation of the spectrum for a particular string vacuum is subject 
also to constraints coming from the RR tadpole
cancellation conditions \cite{lu}. That means the cancellation of
D6-branes charges,\footnote{Taken together with their orientifold
images $(n_a^i, - m_a^i)$ wrapping on three cycles of homology
class~$[\Pi_{\alpha^{\prime}}]$.} wrapping on three cycles with
homology $[\Pi_a]$ and O6-plane 7-form charges wrapping on 3-cycles
with homology $[\Pi_{O_6}]$.  For the toroidal orientifolds we deal in this section, 
the RR tadpole cancellation
conditions in terms of cancellations of RR charges in homology become
\beq
\sum_a N_a [\Pi_a]+\sum_{\alpha^{\prime}} 
N_{\alpha^{\prime}}[\Pi_{\alpha^{\prime}}] - 32
[\Pi_{O_6}]=0\,.
\label{homology}
\eeq  
In explicit form, the RR tadpole conditions read
\beqa
\sum_a N_a n_a^1 n_a^2 n_a^3 = 16\,,
\nonumber\\
\sum_a N_a m_a^1 m_a^2 n_a^3 = 0\,,
\nonumber\\
\sum_a N_a m_a^1 n_a^2 m_a^3 = 0\,,
\nonumber\\
\sum_a N_a n_a^1 m_a^2 m_a^3 = 0\,.
\label{na1}
\eeqa

The complete accommodation of the chiral matter for the five stack SM's can be
seen in table (\ref{spectrum7b}).  
A number of interesting observations are in order:

$\bullet$  There are various low energy gauged symmetries in the
models (as it also happens in the models \cite{iba, kokos2}). 
They are defined in terms of the $U(1)$ symmetries $Q_a$,
$Q_b$, $Q_c$, $Q_d$, $Q_e$ .  Hence the baryon number B is defined as $Q_a =
3B$, the lepton number is $L = Q_d + Q_e$ while $Q_a - 3 Q_d - 3 Q_e =
3 (B-L)$;  $Q_c = 2 I_{3R}$, $I_{3R}$ being the third
component of weak isospin. Also, $3 (B-L)$ and $Q_c$ are free of
triangle anomalies. The $U(1)_b$ symmetry plays the role of a
Peccei-Quinn symmetry in the sense of generating mixed SU(3) anomalies.
From the study of Green-Schwarz mechanism cancellation of anomalies, we
deduce that Baryon and Lepton number are unbroken gauged symmetries
and thus proton should be stable. Also  Majorana masses for right
handed neutrinos are not allowed, meaning that mass terms for
neutrinos should be of Dirac type.

\begin{table}[htb]\footnotesize
\renewcommand{\arraystretch}{3}
\begin{center}
\begin{tabular}{||c||c|c|c|}
\hline
\hline
$N_i$ & $(n_i^1, m_i^1)$ & $(n_i^2, m_i^2)$ & $(n_i^3, m_i^3)$\\
\hline\hline
 $N_a = 3$ & $(1/\b_1, 0)$  &
$(n_a^2, \ {\tilde \epsilon} \epsilon \b_2)$ & $(3, - {\tilde \epsilon} /2)$  \\
\hline
$N_b = 2$  & $(n_b^1, \ \epsilon \b_1)$ & $(1/\b_2, \ 0)$ &
$(1, - {\tilde \epsilon} /2)$ \\
\hline
$N_c = 1$ & $(n_c^1, \ \epsilon {\tilde \epsilon} \b_1 )$ &   $(1/\b_2,\  0)$  &
$(0, \ 1)$ \\
\hline
$N_d = 1$ & $(1/\b_1, \ 0)$ & $(n_d^2,\  2 {\epsilon{\tilde \ep}}\b_2 )$ &  $(1, \  
{\tilde \epsilon}/2)$\\\hline
$N_e = 1$ & $(1/\b_1,\  0)$ & $(n_e^2, {\epsilon{\tilde \ep}}\b_2 )$ &    
$(1, \  {\tilde \epsilon}/2)$
\\\hline
$N_h$ & $(1/ \b^1, 0)$ &
$(1/\b_2, 0 )$
  & $(2, m_h^3)$
\\\hline
\end{tabular}
\end{center}
\caption{\small New general tadpole solutions for the five (5) stack
intersecting
D6-brane model [not appearing in \cite{kokos1}] coming from $IIA/T^6/\Omega R$, giving rise to exactly the
standard model gauge group and observable chiral spectrum
at low energies.
These solutions depend
on two integer parameters,
$n_a^1$, $n_d^1$,
the NS-background $\beta^1= \ 1-b_i$, which is associated to
the presence of the
NS B-field by $b_i =0,\ 1/2$.
 and
the phase parameters $\epsilon = {\tilde \epsilon}=\pm 1$.
\label{spectrum5d}}
\end{table}

 $\bullet$ In order to cancel the appearance of exotic
representations in the model appearing from the $a a^{\star}$ sector,
in antisymmetric and symmetric representations of the $U(N_a)$
group, we will impose the condition
\beq
 {\Pi}_{i=1}^3 m^i = 0\,. 
\label{req1}
\eeq 
The solutions satisfying simultaneously the intersection constraints
and the cancellation of the RR crosscap tadpole constraints are given in 
table~(\ref{spectrum5d}).

 By using the tadpole solutions of table (\ref{spectrum5d})
in~(\ref{na1}) all tadpole equations but the first are 
satisfied, the \footnote{We have added an arbitrary number of $N_h$
branes which don't contribute to the rest of the tadpoles and
intersection number constraints.}  latter becoming :
\beq 
\frac{9 n_a^2}{ \b^1} + 2 \frac{n_b^1}{ \b^2} + \frac{n_d^2}{
\b^1} + \frac{n_e^2}{ \b^1} + N_h \frac{2}{\b^1 \b^2} = 16\,.
\label{ena11}
\eeq 
where we had added the presence of extra $N_h$ branes by setting 
$m_h^3 =0$.  The cancellation of
tadpoles is best described by choosing a consistent numerical set of wrappings,
e.g.\ 
\beq 
n_a^2 = 1, \quad
n_b^1 = 1, \quad
n_c^1 =-1, \quad
n_d^2 = -1, \quad
n_e = -1, \quad
\b^1 = 1/2, \quad
\b^2 = 1, \epsilon {\tilde \epsilon}=1
\label{numero1}
\eeq
With the above choices, all tadpole conditions but the first are
satisfied, the latter is satisfied when we add $N_h = 0 $ extra D$_6$ branes.
  The wrappings become
\beqa
N_a &=& 3\quad (2,  0)(1,   1)(3,  -1/2) 
\nonumber\\
N_b &=& 2\quad (1,   1/2 )(1,  0) (1, -1/2 ) 
\nonumber\\
N_c &=& 1\quad (-1, 1/2)(1, 0)(0, 1) 
\nonumber\\    
N_d &=& 1\quad (2,  0)(-1, 2)(1, 1/2)
\nonumber\\
N_e &=& 1\quad (2, 0) (-1, 1)(1, 1/2 ).
\label{consist}
\eeqa
    
The mixed anomalies $A_{ij}$ of the five $U(1)$'s with the
non-abelian gauge groups $SU(N_a)$ of the theory cancel through a
generalized GS mechanism involving close string modes
couplings to worldsheet gauge fields.  
Three combinations of the
$U(1)$'s are anomalous and become massive.  Two orthogonal
non-anomalous combinations survive massless the Green-Schwarz mechanism, 
one being the hypercharge seen in (\ref{hyper1231}) and also an extra 
U(1) whose value is model dependent and depends on the choice of parameters
seen in table (\ref{spectrum5d}). 
The latter U(1) get broken by demanding that the intersection $ce$ 
respects N=1 supersymmetry, that is
\beq
\pm tan^{(-1)}\left( \frac{\b_1 U^1}{n_c^1}\right) \pm tan^{(-1)}\left( \frac{\b_2 U^2}{n_e^2}
\right) + \frac{\pi}{2}    \pm tan^{(-1)}\left( \frac{U^3}{2}\right) 
\eeq
- where we have chosen $\epsilon {\tilde \epsilon} =1$ -, for some choice of signs.
In the special case, $n_e^2 = 0$, we get the N=1 susy condition 
\beq
\frac{\b_1 U^1}{n_c^1}= \frac{U^3}{2}
\eeq
constraining the complex structure parameters $U^1$, $U^3$. The choice $n_c^1 =1$ determines
the fifth U(1) to be $F_a + (28/3) F_c - F_e$. We note that on this classes of models  the string scale
cannot be lowered at the TeV region according to the geometric scenario of \cite{dva} and 
thus the solution of gauge hierarchy problem in the Higgs sector is an open question.
Whether of not, the supersymmetry present in particular sectors of the theory, or additional 
superymmetry - that can be implemented in the models by some choice of parameters - could be of 
any help remains to be seen.     

The 6-stack SM configuration that was localized in toroidal orientifold 
compactifications from intersecting D6-branes in \cite{kokos2} was given in
table (\ref{spectrum7c}). Here we will present an alternative solution to the 
RR tadpoles. The new solution can be seen in table (\ref{kouno}). We have 
also added an arbitrary number of extra $N_h$ branes that might needed in order
to satisfy the RR tadpoles. For these solutions the procedures followed 
in \cite{kokos2} could be also repeated to examine further the SM's.

\begin{table}[htb]\footnotesize
\renewcommand{\arraystretch}{3}
\begin{center}
\begin{tabular}{||c||c|c|c|}
\hline
\hline
$N_i$ & $(n_i^1, m_i^1)$ & $(n_i^2, m_i^2)$ & $(n_i^3, m_i^3)$\\
\hline\hline
 $N_a = 3$ & $(1/\b_1, 0)$  &
$(n_a^2, \ {\tilde \epsilon} \epsilon \b_2)$ & $(3, - {\tilde \epsilon} /2)$  \\
\hline
$N_b = 2$  & $(n_b^1, \ \epsilon \b_1)$ & $(1/\b_2, \ 0)$ &
$(1, - {\tilde \epsilon} /2)$ \\
\hline
$N_c = 1$ & $(n_c^1, \ \epsilon {\tilde \epsilon} \b_1 )$ &   $(1/\b_2,\  0)$  &
$(0, \ 1)$ \\
\hline
$N_d = 1$ & $(1/\b_1, \ 0)$ & $(n_d^2,\   {\epsilon{\tilde \ep}}\b_2 )$ &  
$(1, \  {\tilde \epsilon}/2)$\\\hline
$N_e = 1$ & $(1/\b_1,\  0)$ & $(n_e^2, {\epsilon{\tilde \ep}}\b_2 )$ &    
$(1, \  {\tilde \epsilon}/2)$
\\\hline
$N_f = 1$ & $(1/\b_1,\  0)$ & $(n_f^2, {\epsilon{\tilde \ep}}\b_2 )$ &    
$(1, \  {\tilde \epsilon}/2)$
\\\hline
$N_h$ & $(1/ \b^1, 0)$ &
$(1/\b_2, 0 )$
  & $(2, m_h^3)$
\\\hline
\end{tabular}
\end{center}
\caption{\small New general tadpole solutions for the six (6) stack
intersecting
D6-brane model [not appearing in \cite{kokos2}] coming from $IIA/T^6/\Omega R$, giving rise to exactly the
standard model gauge group and observable chiral spectrum
at low energies.
These solutions depend
on two integer parameters,
$n_a^1$, $n_d^1$,
the NS-background $\beta^1= \ 1-b_i$, which is associated to
the presence of the
NS B-field by $b_i =0,\ 1/2$.
 and
the phase parameters $\epsilon = {\tilde \epsilon}=\pm 1$.
\label{kouno}}
\end{table}

\section{Building the $SU(4)_C \times SU(2)_L \times SU(2)_R$ GUTS With Only
The SM At Low Energy} 

Extensions of these GUTS with four, five and six stacks of D6-branes were also considered in \cite{kokos5, kokos6, kokos7}. 
\newline
The basic features found in these intersecting 
D6-brane models can be classified as 
follows : \newline 
$\bullet$ The models even though they have overall N=0 SUSY,
possess N=1 SUSY subsectors which are necessary in order to create a 
Majorana mass term for $\nu_R$'s. \newline 
$\bullet$ Extra branes are needed to cancel RR tadpoles. The presence of 
these
branes creates extra matter singlets, transforming under both the visible SM
gauge group and the extra D6-brane gauge group that may be 
used to break the extra U(1)'s, beyond hypercharge, surviving massless the 
presence of the generalized Green-Schwarz mechanism. Their presence is also 
used to make massive the exotic fermions, seen for example in the 
bottom part of table (1), taken from \cite{kokos7}.  
\begin{table}[htb] \footnotesize
\renewcommand{\arraystretch}{0.8}
\begin{center}
\begin{tabular}{|c|c||c|c||c||c|c|c|}
\hline
Fields &Intersection  & $\bullet$ $SU(4)_C \times SU(2)_L \times SU(2)_R$
 $\bullet$&
$Q_a$ & $Q_b$ & $Q_c$ & $Q_d$ & $Q_e$\\
\hline
 $F_L$& $I_{ab^{\ast}}=3$ &
$3 \times (4,  2, 1)$ & $1$ & $1$ & $0$ &$0$ &$0$\\
 ${\bar F}_R$  &$I_{a c}=-3 $ & $3 \times ({\ov 4}, 1, 2)$ &
$-1$ & $0$ & $1$ & $0$ & $0$\\
 $\chi_L^1$& $I_{bd^{\star}} = -8$ &  $8 \times (1, {\ov 2}, 1)$ &
$0$ & $-1$ & $0$ & $-1$ & $0$\\    
 $\chi_R^1$& $I_{cd} = -8$ &  $8 \times (1, 1, {\ov 2})$ &
$0$ & $0$ & $-1$ &$1$ &$0$\\
 $\chi_L^2$& $I_{be} = -4$ &  $4 \times (1, {\ov 2}, 1)$ &
$0$ & $-1$ & $0$ &$0$ & $1$ \\    
 $\chi_R^2$& $I_{ce^{\ast}} = -4$ &  $4 \times (1, 1, {\ov 2})$ &
$0$ & $0$ & $-1$ & $0$ &$-1$ \\\hline
 $\omega_L$& $I_{aa^{\ast}}$ &  $6 \b^2
 \times (6, 1, 1)$ & $2$ & $0$ & $0$ &$0$ &$0$\\
 $y_R$& $I_{aa^{\ast}}$ & $6  \b^2  \times ({\bar 10}, 1, 1)$ &
$-2$ & $0$ & $0$ &$0$ &$0$ \\
\hline
 $s_R^1$ & $I_{dd^{\ast}}$ &  $16 \b^2
 \times (1, 1, 1)$ & $0$ & $0$ & $0$ &$2$ &$0$\\
 $s_R^2$ & $I_{ee^{\ast}}$ & $8  \b^2  \times (1, 1, 1)$ &
$0$ & $0$ & $0$ &$0$ &$-2$ \\
\hline
\end{tabular}
\end{center}
\caption{\small Fermionic spectrum of the $SU(4)_C \times
SU(2)_L \times SU(2)_R$, PS-II class of models together with $U(1)$
charges. We note that at energies of order $M_z$ only
the Standard model survives.
\label{spectrum8}}
\end{table}
The fermion spectrum of table (1) is consistent with the calculation of RR
tadpoles. The RR tadpoles get cancelled with the introduction of extra U(1) 
branes, $h^i$, that transform under the both the extra U(1) gauge group 
and the 
rest of the
intersecting D6-branes of table (1). The existence of N=1 SUSY at 
the intersections $dd^{\star}$, $dh$, $dh^{\star}$, $eh$, $eh^{\star}$,   creates the singlets $s_1^B$, $\kappa_3^b$, $\kappa_4^b$, 
$\kappa_5^b$, $\kappa_6^b$ respectively, that contribute to the mass of the
`light' fermions $\chi_L^1$, $\chi_L^2$. All fermions of table (1) receive 
a mass of order $M_s$; the only exception being the light masses of
$\chi_L^1$, $\chi_L^2$, weak fermion doublets. 
Lets us discuss the latter issue in more detail.
\newline
The left handed fermions $\chi_L^1$ receive a contribution to their mass
from the coupling \footnote{
In (\ref{ka1sa1}) we have included the leading contribution of the 
worksheet area connecting the
seven vertices. In the following for simplicity reasons we will set the 
leading contribution of the different couplings to
one e.g. area tends to zero.}
\beq
(1, 2, 1)(1, 2, 1) e^{-A}
 \frac{\langle h_2 \rangle \langle h_2 \rangle
\langle {\bar F}_R^H  \rangle \langle H_1 \rangle
\langle {\bar s}_B^1 \rangle}{M_s^4}
\stackrel{A \rightarrow 0}{\sim}
\frac{\upsilon^2}{M_s} \ (1, 2, 1)(1, 2, 1)
\label{ka1sa1}
\eeq 
and from another coupling, of the same order as (\ref{ka1sa1}), also
contributing to the mass of the $\chi_L^1$ fermion as 
\begin{eqnarray}
(1, 2, 1)(1, 2, 1) \frac{ \langle h_2 \rangle \langle h_2 \rangle
\langle {\bar F}_R^H \rangle \langle H_1 \rangle
 \langle {\bar \kappa}_3^B \rangle
 \langle {\bar \kappa}_4^B \rangle }{M_S^9} \sim
(1, 2, 1)(1, 2, 1)\frac{\upsilon^2}{M_s},
\label{addi1}
\end{eqnarray}
The left handed fermions $\chi_L^2$
receives a non-zero mass
from the coupling 
\beq
(1, 2, 1)(1, 2, 1) 
 \frac{\langle h_2 \rangle \langle h_2 \rangle
\langle {\bar F}_R^H  \rangle \langle H_1 \rangle
\langle  {\bar s}_B^2 \rangle}{M_s^4}
\stackrel{A \rightarrow 0}{\sim}
\frac{\upsilon^2}{M_s} \ (1, 2, 1)(1, 2, 1)
\label{ka1sa2}
\eeq
and the coupling
\begin{eqnarray}
(1, 2, 1)(1, 2, 1) \frac{ \langle h_2 \rangle \langle h_2 \rangle
\langle {\bar F}_R^H \rangle \langle H_1 \rangle
 \langle {\bar \kappa}_5^B \rangle
 \langle {\bar \kappa}_6^B \rangle }{M_S^5} \sim
(1, 2, 1)(1, 2, 1)\frac{\upsilon^2}{M_s},
\label{addi2}
\end{eqnarray}
Thus assuming that the leading area Yukawa for
the couplings is of order ${\cal O}(1)$, e.g. associated areas going to zero,
the masses of \footnote{In this case the masses of $\chi_L^1$, $\chi_L^2$ are the sum of t
he contributions
of (\ref{ka1sa1}, \ref{addi1}) and (\ref{ka1sa2}, \ref{addi2}) respectively
}
\begin{equation}
\chi_L^1,\  \chi_L^2 \sim \frac{2 \upsilon^2}{M_s} \,
\end{equation}
$\bullet$ As the particles $\chi_L^1,\  \chi_L^2$ are not observed at present,
the fact that their mass may be between 
\beq
100 \ GeV \ \ \leq \ \chi_L^1, \ \chi_L^1 \  \leq \  2 \upsilon = \
max \{ \frac{2\upsilon^2}{M_s}\} =\ 492 \ GeV
\eeq
sends the string scale 
\beq
M_s  \leq 1.2 \ TeV
\eeq
This is a general feature of all the Pati-Salam models based 
on toroidal orientifolds; they 
 predict the existence of light weak doublets with masses \footnote{The reader may convince
 itself that the maximum value of $2 \upsilon^2 / M_s$ is $2 \upsilon$.}  
 between 100 and
 $\upsilon = 492$ GeV.
 The latter result may be considered
 as a general
 prediction of all classes of models based on intersecting
 D6-brane Pati-Salam GUTS.
  \newline 
 Another important property of these constructions is that
the conditions for some intersections to respect N=1 supersymmetry and also
needed to guarantee the existence of a Majorana mass term for s$\nu_R$'s : 
\newline  
$\bullet$ solve the orthogonality conditions for the extra - beyond 
hypercharge - U(1)'s \footnote{The latter becoming massive from the use of 
extra singlets created by the presence of extra branes; the latter
 needed to satisfy the 
RR tadpoles.} to survive massless the presence of a generalized
Green-Schwarz mechanism describing the couplings of the U(1)'s
to the RR two form fields. \newline   
The considerations we have just described \cite{kokos5}, \cite{kokos6}, 
\cite{kokos7} are quite generic and 
the same methodology
applies easily
to the construction of more general GUT gauge groups in the context of
intersecting brane worlds.

We note that at present the only existing string GUT constructions,
in the context of Intersecting D6-brane Models, 
that have only the SM at low energy, with complete cancellation 
of RR tadpoles, are: \newline 
a) the toroidal orientifold II Pati-Salam GUTS of 
\cite{kokos5, kokos6, kokos7} and 
\newline
b) the constructions of flipped SU(5), and 
SU(5) GUTS of \cite{kokos4} described next.


\section{The Construction of SU(5), Flipped SU(5) GUTS with only
the SM at Low Energy} 

Lets us review the intersecting D6-branes constructions of 
the $Z_3$ orientifolds of \cite{lust3}. The D6-branes involved satisfy
the following RR tadpole conditions where 
\footnote{
The net number of bifundamental massless chiral fermions in the models
is defined as
\beqa
({\bar N}_a, N_b)_L :\  I_{ab} = Z_a Y_b - Y_a Z_b \\
(N_a, N_b)_L : \   I_{ab^{\star}} = Z_a Y_b + Y_a Z_b
\label{spec1}
\eeqa
}
\beq
\sum_a N_a Z_a = 2 \ .
\label{tad}
 \eeq
As it was noticed in \cite{lust3} the simplest realization of an SU(5) GUT
involves two stacks of D6-branes at the string scale $M_s$, the
first one corresponding to a $U(5)$ gauge group while the second one to a
U(1) gauge group. Its effective wrapping numbers are given by 
\beq (Y_a,
Z_a) = (3, \frac{1}{2}), \  (Y_b, Z_b) = (3, -\frac{1}{2}),
\label{ena} \eeq Under the decomposition $ U(5) \subset SU(5)
\times U(1)_a$, the models become effectively an $SU(5) \times
U(1)_a \otimes U(1)_b$ GUT. One combination of U(1)'s
become massive due to its coupling to a RR field, 
another one remains massless to low energies. The spectrum of this
SU(5) GUT may be seen in the first seven columns (reading from the left) 
of table (2).
\begin{table}[htb]\footnotesize
\renewcommand{\arraystretch}{1.5}
\begin{center}
\begin{tabular}{||c|c|c|c|c|c|c|c||}
\hline \hline
Field & Sector & Multip. & $SU(5)$    &  $U(1)_a$ & $U(1)_b$
 & $U(1)^{mass}$ & $U(1)^{fl} = \frac{5}{2} \times U(1)^{mass} $\\\hline
$f$& \{ 51 \} & $3$          & ${\bf {\bar 5}}$ & $-1$      &  $1$   & $-\frac{6}{5}$ & -3    \\\hline
$F$ & $A_{a}$ & $3$          & ${\bf 10}$   &  2         &   0      &$\frac{2}{5}$ &   1          \\\hline
$l^c$ & $S_{b}$ & $3$          &  ${\bf 1} $         &   0        &   -2   & $2$ & 5    \\\hline
  \hline
\end{tabular}
\end{center}
\caption{Chiral Spectrum of a two intersecting D6-brane stacks in a three 
generation flipped $SU(5) \otimes U(1)^{mass}$ model.
Note that the charges under the $U(1)^{fl}$ gauge symmetry,
 when rescaled
appropriately (and $U(1)^{fl}$ gets broken) `converts' the flipped SU(5) model
to a three
generation (3G) $SU(5)$. 
\label{flip}}
\end{table}
At this stage the SU(5) models - have the correct chiral 
fermion content of an SU(5) GUT - and the extra U(1) surviving 
the presence of the Green-Schwarz mechanism, 
breaks by the use of a singlet field present. However, 
the electroweak ${\bf }5$-plets needed for electroweak
 symmetry breaking of the models are absent. Later on, attempts to 
construct a
fully N=1 supersymmetric SU(5) models at $M_s$ in \cite{cve4}, produced 
3G models
that were not free of remaining massless exotic 15-plets. 
Also, later on 
in \cite{nano} it was noticed that if one leaves unbroken, 
and rescales, the U(1) 
surviving massless the Green-Schwarz mechanism of the SU(5) GUT of 
\cite{lust3}, the rescaled U(1) becomes
the flipped U(1) generator. However, the proposed 3G models lack the presence
of GUT Higgses or electroweak pentaplets and were accompanied by extra
exotic massless matter to low energies.

In \cite{kokos4} we have shown that it is possible to construct the first 
examples of string 
SU(5) and
 flipped
SU(5) GUTS - where we identified the appropriate GUT and electroweak 
Higgses - which break to the SM at low energy.
E.g. in the flipped SU(5) GUT, the fifteen fermions of the SM plus the right
handed neutrino $\nu^c$ belong to the 
\beq
F = {\bf 10_1} = (u, d, d^c, \nu^c), \ \
f = {\bf {\bar 5}_{-3}} = (u^c, \nu, e), \  \ l^c =  {\bf 1_{5}} = e^c
\label{def1}
\eeq
chiral multiplets.
The GUT breaking Higgses may come from the `massive' spectrum
of the sector localizing
the ${\bf 10}$-plet (${\bf 10_1^B} = (u_H, d_H, d^c_H, \nu^c_H )$ 
fermions seen in table (\ref{flip}). The lowest order
Higgs in this sector, let us call them $H_1$, $H_2$, have quantum numbers as
those given in table (\ref{Higgsfli}).
\begin{table} [htb] \footnotesize
\renewcommand{\arraystretch}{1}
\begin{center}
\begin{tabular}{||c|c||c|c|c|c||}
\hline
\hline
Intersection & GUT Higgses & repr. & $Q_a$ & $Q_b$ & $Q^{fl}$\\
\hline\hline
$\{ a,{\tilde O6} \}$  & $H_1$  &  {\bf 10}   & $2$   & $0$ & $1$ \\
\hline
$ \{ a,{\tilde O6} \}$  & $H_2$  &  ${\bf {\bar 10}}$   & $-2$ & $0$  & $-1$ \\
\hline
\hline
\end{tabular}
\end{center}
\caption{\small 
Flipped $SU(5) \otimes U^{fl}$ GUT symmetry breaking
scalars. 
\label{Higgsfli}}
\end{table}
By looking at the last column of table (\ref{Higgsfli}), we realize that the
Higgs $H_1$, $H_2$ are the GUT symmetry breaking Higgses of a standard
flipped SU(5) GUT.
By dublicating the analysis of section (3.1),
one may conclude that what it appears in the effective theory as GUT 
breaking Higgs scalars, is the combination 
$
H^G = H_1 + H_2^{\star}$.
In a similar way the correct identification
of the electroweak content \cite{kokos4} of the flipped SU(5) ${\bf 5_{-2}^B} = (D, h^{-},  h^{0} )$-plet
(and 
SU(5))GUTS made possible 
the existence of the see-saw mechanism which is generated by the 
interaction \beqa {\cal
L} = {\tilde Y}^{\nu_L \nu_R} \cdot {\bf 10} \cdot {\bar {\bf
5}} \cdot {\bar h}_4 \
      + \  {\tilde Y}^{\nu_R} \cdot \frac{1}{M_s} \cdot ({\bf 10} \cdot
{\bf \overline{10}}^B)
      ({\bf 10} \cdot {\bf \overline{10}}^B)\ . 
\label{seesaw1}
\eeqa
Its standard version can be generated by choosing
\beq
\langle h_4 \rangle = \upsilon, \  \langle {\bf 10}_i^B \rangle = M_s
\label{masse1}
\eeq
and generates small neutrino masses.
In these constructions the baryon number is not a gauged symmetry, thus 
a high GUT scale of the order of the $10^{16}$ GeV helps the theory to
avoid gauge mediated proton decay modes like the \cite{kokos4}
\beq
  \sim \frac{1}{M_s^2}\ ({\bar u}^c_L \ u_L) \ ({\bar e}_{R}^{+}) 
(d_{R}),\ \
\sim \frac{1}{M_s^2}\ ({\bar d}^c_R \ u_R) ({\bar d}^c_L \ \nu_L) \ .
\eeq   
[In IBW's proton decay by direct calculation of string amplitudes
for SUSY SU(5) D-brane models was examined in \cite{igorwi}.]
Scalar mediated proton decay modes get suppressed by the existence of 
a new solution to the doublet-triplet splitting problem   
 \beq
\frac{r}{M_s^3}(HHh)( {\bar F}  {\bar F} {\bar h}) + m ({\bar h}h) (
{\bar H} H) + \kappa ({\bar H}H)( 
{\bar H} H),\eeq
that stabilizes the vev's of the triplet scalars $d_c^H$, $D$ \cite{kokos4}.
This is the first example of a doublet-triplet splitting realization
in IBW's. 
The full solution of the gauge hierarchy problem,
that is avoiding the existence of quadratic corrections to the electroweak 
Higgses remains an open issue in the present GUTS.  

 Recently the interest of model building in IBW's has been focused in 
the 
construction of non-susy intersecting D6-brane models which localize the 
spectrum of MSSM at low energies
\cite{ibanew}, \cite{kokosnew}.


\section{The SM at low energy from $Z_3$, $Z_3 \times Z_3$ orientifolds }

The SM but with some of the matter not in bifundamnetal reps have been derived intially
in \cite{lust3} in the context of$Z_3$  orientifolds. Quite recently \cite{kokosnew}, we have shown
that using  $Z_3 \times Z_3$  orientifolds, we can reproduce these SM vacua 
for various choices of wrappings. In this section, we present some novel features of the new 
constructions, including the derivation of the SM. 

In $Z_3$, $Z_3 \times Z_3$ orientifolds all complex structure moduli are fixed from the 
beginning. In particular the $Z_3 \times Z_3$ orientifold models are subject to the cancellation of untwisted RR tadpole 
conditions \cite{kokosnew} given by
 \beq
\sum_a N_a  Z_a = 4, 
\label{tad}
\eeq    
where
\beqa
Z_a = 2 m^1_a m^2_a m^3_a + 2 n^1_a n^2_a n^3_a - n_a^1 n_a^2 m_a^3 - 
n_a^1 m_a^2 n_a^3 - 
m_a^1 n_a^2 n_a^3 - \nonumber\\
m_a^1 m_a^2 n_a^3 -m_a^1 n_a^2 m_a^3 - n_a^1 m_a^2 m_a^3 \ .
\eeqa
The gauge group U($N_a$) supported by $N_a$ coincident D6$_a$-branes comes 
from the $a({\cal O}a)$ sector, the sector made from 
open strings streched between the $a$-brane and its images under the orbifold 
action. In addition, we get three adjoint N=1 chiral multiplets. 
The $a({\cal O}b)$ sector, strings stretched between the brane $a$ and the 
orbit images of brane $b$, will give $I_{ab}$ fermions in the 
bifundamental $(N_a, {\bar N}_b)$ where 
\beq
I_{ab}= 3(Z_a Y_b - Z_b Y_a),
\eeq 
and $(Z, Y) $ are the effective wrapping numbers with $Y$ given by
\beq
Y_a = m^1_a m^2_a m^3_a + n^1_a n^2_a n^3_a - n^1_a n^2_a m^3_a - 
n^1_a m^2_a n^3_a - m^1_a n^2_a n^3_a 
\eeq
The sign of $I_{ab}$ denotes the chirality of the associated fermion, where
we choose positive intersection numbers for left handed fermions. 
In the sector ${ab^{\prime}}$ - strings stretching between the brane $a$ and
the orbit images of brane $b$, there are $I_{ab^{\prime}}$ chiral fermions 
in the bifundamental $(N_a, N_b)$, with
\beq
I_{ab^{\prime}}= 3(Z_a  Z_b - Z_a Y_b - Z_b Y_a),
\eeq 
The following numbers of chiral fermions
in symmetric (S) and antisymmetric (A) representations of $U(N_a)$, open
strings stretching between the brane $a$ and its orbit images $({\cal O}a)$, 
are also included 
\beqa
(A_a) = 3(Z_a - 2 Y_a) ,\\
(A_a + S_a) = \frac{3}{2}(Z_a - 2 Y_a)(Z_a - 1)
\eeqa 
Also, from open strings streched between the brane $a$ and its orbifold 
images we accommodate non-chiral massless fermions in the adjoint representation,
\beq
(Adj)_L : \prod_{i=1}^3 (L^I_{[a]})^2 \ ,
\label{adj1}
\eeq
where 
\beq
L^I_{[a]} = \sqrt{(m_a^I)^2 + (n_a^I)^2 -(m_a^I) (n_a^I) }
\eeq 
Adjoint massless matter, including fermions and gauginos, 
is expected to become massive from loops once supersymmetry is broken leaving only the gauge bosons massless.  
Supersymmetry may be preserved by a system of branes if each stack of 
D6-branes is related to the $O6$-planes by a rotation in $SU(3)$, that
is the angles ${\tilde \theta}_i $ of the D6-branes with respect to 
the horizontal
direction in the i-th two-torus obeys the condition ${\tilde \theta}_1 + 
{\tilde\theta}_2 + {\tilde \theta}_3 = 0$. 
In the low energy theory, cubic gauge anomalies automatically cancel, 
due to the RR tadpole conditions (\ref{tad}). Mixed U(1)-gauge anomalies 
also cancel due to
the existence of a generalized Green-Schwarz mechanism 
 that makes massive only one U(1) gauge field given by
\beq
\sum_a N_a (Z_a - 2 Y_a) F_a
\label{masi}
\eeq

By choosing to work with three stacks of intersecting D6-branes and making the choice of wrapping numbers
\beq
(Z_a, Y_a) = \left( \ba{cc} 1, & 0 \\\ea \right), \ (Z_b, Y_b) = \left(\ba{cc}1, 
&  1 \\\ea \right), \ (Z_c, Y_c) = \left( \ba{cc} -1, & -1 \\\ea \right)
\label{wrap341}
\eeq 
we find exactly the 
(non-supersymmetric) SM spectrum that may be seen
in table (\ref{tabold}).

\begin{table}[htb] \footnotesize
\renewcommand{\arraystretch}{2}
\begin{center}
\begin{tabular}{|r|c|c|}
\hline\hline
 ${\bf Matter\hspace{2cm}}$ & $(SU(3) \times SU(2))_{(Q_a, Q_b, Q_c)} \hspace{2cm}$ & $U(1)^{Y}$ \\
\hline\hline
$\hspace{2cm} \{ Q_L \} \hspace{2cm}$   & $3(3, {2})_{(1,\ -1,\ 0)} \hspace{2cm}$ & $1/6$  \\
\hline
$\hspace{2cm}\{ u_L^c \}\hspace{2cm}$   & $3({\bar 3}, 1)_{(2,\ 0,\ 0)}\hspace{2cm}$ & $-2/3$  \\
\hline
$\hspace{2cm}\{ d_L^c \}\hspace{2cm}$ & $3(3, 1)_{(-1,\ 0, \  1)}\hspace{2cm}$ & $1/3$ \\
\hline
$\hspace{2cm}\{ L \}\hspace{2cm}$ & $3(1, 2)_{(0,\  1,\  1 )}\hspace{2cm}$ & $-1/2$ \\
\hline
$\hspace{2cm}\{ e_L^{+} \}\hspace{2cm}$ & $3(1, 1)_{(0, \ -2,\ 0)}\hspace{2cm}$ & $1$ \\
\hline
$\hspace{2cm}\{ N_R \}\hspace{2cm}$ & $3(1, 1)_{(0, \ 0, \ -2 )}\hspace{2cm}$ & $0$ \\
\hline
\end{tabular}
\end{center}
\caption{\small
A three generation chiral (open string) spectrum accommodating the SM. 
The required Higgs may come from 
 bifundamental N=2 hypermultiplets in the N=2 $bc$, $bc^{\star}$ 
sectors \cite{iba, kokos1, kokos2} that may trigger brane recombination.
\label{tabold}}
\end{table}
This is exactly the SM found in \cite{lust3}. It can be reproduced \cite{kokosnew}
with many choices of effective wrappings $(Z, Y)$.
As we can see, many of the matter fields present are not in bifundamental representations.
In this case, we find that there is no mass term for the up-quarks allowed by charge 
conservation, which is not the case for the SM's of section 3 and \cite{iba, kokos1, kokos2}. 

\section{Conclusions}

IBW's offer a lot of promising research directions. We note that one of the major problems  
in string model building, and in string theory in general, is the determination of the values of the
moduli. As moduli are free parameters, the determination of the single 
vacuum in string theory requires that there is a dynamical determination 
of their value.
IBW's offer a solution towards this direction, as we have seen in section 3, 
 as the presence of N=1 supersymmetry in particular sectors of the theory 
can fix the complex structure moduli. On the other hand SM's coming from
$Z_3 \times Z_3$ orientifolds have from the start all complex structure 
moduli fixed. In this respect, $Z_3 \times Z_3$ orientifolds solve 
in principle the problem of determining complex structure moduli. Thus 
model building attempts coming from the latter constructions offer
 a promising 
avenue for further studies as it only remains to show that it is 
possible that also the K\"ahler moduli can be also fixed (possibly from
non-perturbative effects). 
We also note that the phenomenology of models with 
intersecting D5-branes on 
$IIB/\Omega R/(T^4 \times C/Z_N)$ (of section 2) need to be further studied as these 
Standard models suffer from no gauge hierarchy problem while they are 
also consistent with the extra dimension scenario \cite{dva}.

\end{document}